\documentstyle[12pt]{article}
\newcommand{\beq}{\begin{equation}}
\newcommand{\eeq}{\end{equation}}
\newcommand{\beqa}{\begin{eqnarray}}
\newcommand{\eeqa}{\end{eqnarray}}
\def\sm{Standard Model}
\def\la{~\mbox{\raisebox{-.6ex}{$\stackrel{<}{\sim}$}}~}
\def\ga{~\mbox{\raisebox{-.6ex}{$\stackrel{>}{\sim}$}}~}
\def\tl{$\tilde t_L$}
\def\tr{$\tilde t_R$}

\def\npb#1{Nucl.\ Phys.\ {\bf B #1}}
\def\plb#1{Phys.\ Lett.\ {\bf B #1}}
\def\prd#1{Phys.\ Rev.\ {\bf D #1}}
\def\prl#1{Phys.\ Rev.\ Lett. {\bf #1}}

\def\ijmp#1{Int.\ J.\ Mod.\ Phys.\ {\bf A #1}}

\def\arnp#1{{Ann.~Rev.~Nucl.~Part.~Sci.~}{\bf #1}}
\begin{document}

\title{Baryogenesis and Low Energy $CP$ Violation}

\author{Mihir P. Worah \\
Department of Physics \\
University of California, Berkeley, California 94720 \\
and \\
Theoretical Physics Group \\
Lawrence Berkeley National Laboratory}

\maketitle

\vspace{-265pt}
\begin{center}
  \hfill    LBNL-44150 \\
~{}  \hfill    August 1999 \\
\end{center}
\vspace{200pt}

\begin{abstract}

$CP$ violation is a crucial component in the creation of the matter -
anti matter asymmetry of the universe. An important open question is 
whether the $CP$ violating phenomena observeable in terrestrial experiments 
have any relation with those responsible for baryogenesis.
We discuss two mechanisms of baryogenesis where this question can
be meaningfully posed: ``electroweak baryogenesis'' and ``baryogenesis 
via leptogenesis''. We show how these scenarios can be constrained by
existing and forthcoming experimental data. We present a specific
example of both these scenarios where the $CP$ violating phase in the
Cabbibo Kobayashi Maskawa matrix is related in a calculable way to the 
$CP$ violating phase responsible for baryogenesis. 

\end{abstract}

The world that we observe is manifestly baryon asymmetric. All the
stable matter we see is made up of baryons,
with anti-baryons being created
only in high energy collisions (either in the laboratories or out in
the cosmos). There is evidence that this asymmetry persists even at
much larger scales.
Matter and anti-matter galaxies within the same galactic cluster
would result in strong $\gamma$ ray emission due to annihilations. The
absence of these confirms a baryon asymmetric region on the 20 Mpc scale
\cite{steigman}. More recently, a bound on the scale of the observable
universe has been obtained by ruling out a contribution to the diffuse
$\gamma$ ray spectrum from particle-antiparticle annihilation
\cite{CDG}. The observed nuclear abundances in the stars, then allows
us to estimate that the current baryon to photon ratio, $n_B/n_{\gamma}
= (4-7)\times 10^{-10}$. This corresponds to a baryon-antibaryon
asymmetry of 1 part in $10^8$ in the early universe.

One possible explanation for the asymmetry is that it is an initial
condition, that we cannot hope to understand. The other,
more appealing, possibility is that although the universe initially
had no net baryon number, microphysical processes that we can hope to
understand led it to develop one during its evolution from the big
bang to the present epoch. There are three requirements in order for
such a baryon asymmetry to develop \cite{sakharov}:

\begin{description}
\item{($i$)}
{There must be a departure from thermal equilibrium. $CPT$ invariance
gaurantees the equality of particle and anti-particle masses. Hence in
thermal equilibrium both will have the same number density as dictated
by Boltzmann statistics.}

\item{($ii$)}
{There must be baryon number violation. This requirement is self
explanatory.}

\item{($iii$)}
{There must be $C$ and $CP$ violation. 
This is required in order for the above
baryon number violating interactions to preferentially produce baryons
rather than antibaryons.}
\end{description}
The discovery of $CP$ violation in the neutral $K$ mesons, thus 
made possible a meaningful discussion, in terms of physical processes, 
of why the universe consists of only matter and no anti-matter.

It was later realised that 
the \sm, in fact, contains all of the three ingredients listed
above that are required for baryogenesis \cite{KRS}.  
At a temperature $T \sim 100$ GeV in the early universe, 
the electroweak symmetry was
broken due to the Higgs field aqquiring a vacuum expectation
value. This resulted in a phase transition
which, if strong enough, could provide the departure from thermal
equilibrium needed for baryogenesis. 
Although baryon number is conserved in the \sm\ at the classical
level, it is broken at the quantum level due to the anomolous coupling 
of the $B + L$ (baryon number plus lepton number) 
current to two $W$ bosons. This baryon
number violation is unobservably small at zero temperature, but it is
enhanced at high temperatures, and could be a viable source for the
asymmetric creation of baryons over anti-baryons. Finally, $CP$
violation has been observed in the neutral $K$'s, and is explained by
a complex phase in the Cabbibo, Kobayashi, Maskawa (CKM) matrix.
 
Unfortunately, in the \sm, neither is the phase transition strong
enough, nor is the $CP$ violation efficient enough to explain the
observed baryon asymmetry. 
The requirement on the strength of the phase transition in order to be 
able to generate and mantain a baryon asymmetry is given by
\cite{Shaposhnikov1}
\beq
\frac{H(T_0)}{T_0} \ge 1.
\label{firstorder}
\eeq
Here $T_0 \sim 100$ GeV 
is the critical temperature for the phase transition, and
$H(T_0)$ is the value of the Higgs vacuum expectation value at this
temperature. This strength is governed by the ratios of boson masses 
that are generated by the spontaneous symmetry breaking (SSB) 
to the mass of the Higgs boson. In the \sm, the $W$ and $Z$
bosons get their masses by SSB, 
and one obtains the approximate relationship
\beq
\frac{H(T_0)}{T_0} \sim \frac{2 M_W^3 + M_Z^3}{2 m_H^2 v} 
                   \sim \frac{1}{2}.
\label{smtransition}
\eeq
Where we have used $M_W = 80$ GeV, $M_Z = 90 GeV$, $m_H = 95 GeV$ 
(which is the current LEP lower bound), 
and $v = 246$ GeV is the zero temperature Higgs vacuum expectation value. 

Assuming a strong enough phase transition and perfectly efficient
baryon number violation one can obtain the estimate 
$n_B/n_{\gamma} \sim 10^{-2} \delta$
where $\delta$ is a dimensionless measure of $CP$
violation \cite{dine}.                          
However, the CKM mecahnism of $CP$ violation in the \sm\ requires the 
participation of all three fermion families, and $\delta$ will be
proportional to $Det~{\cal C}/T_0^{12} \sim 10^{-21}$, where 
$Det~{\cal C}$ is the Jarlskog determinant \cite{Jarlskog}, 
and we have used $T_0 = 100$ GeV.
There is a further suppression since
the time scale needed for such interactions is so large that 
finite temperature plasma effects cause the
participating particle wave functions to decohere before they can
interfere enough to generate a significant $CP$ asymmetry \cite{HS}. 
Thus, it is clear that one needs to invoke physics beyond the \sm\ in
order to explain the baryon asymmetry of the universe. 

In this talk we 
give an overview of baryogenesis in two extensions of the \sm. 
These models are motivated by the fact that they offer explanations
for observed phenomena other than the baryon asymmetry that cannot be
explained by the \sm\ and, most importantly, have low energy
experimental consequences. We will demonstrate that
in these models it is possible to
relate the $CP$ violation responsible for baryogenesis with the $CP$
violation observed in the neutral Kaons. 

One obvious possibility is to augment the \sm\ with new particles in
the 100 GeV mass range that would remedy the deficiencies pointed out
above \cite{EWB}. 
Additional bosons that get their masses by the Higgs mechanism
could enhance the strength of the electroweak phase
transition. Moreover, the 
richer particle content could make $CP$ violation more efficient.
The most attractive such extension is the Minimal Supersymmetric
Standard Model (MSSM), which we consider here. 
This model has its primary motivations in the facts that
it stabilizes the hierarchy between the electroweak scale and the
Planck scale, and that it provides a natural explanation of electro
weak symmetry breaking. 

The other distinct possibility is to use the 
baryon and/or lepton number, and $CP$ violating decays of some
super-heavy particle. The departure from thermal equilibrium typically
occurs because the decay rate of the particle is slower than the
expansion rate of the universe \cite{olive}.
These processes must occur in the very early history of the universe
because it is only then that the expansion rate was rapid enough to
provide the out of equilibrium conditions needed for baryogenesis.
The situation we will consider is where the \sm\ is 
augmented with massive ($\sim 10^{10}$ GeV) right-handed
Majorana neutrinos which have lepton number and $CP$ violating mass
matrices. Their out of equilibrium decays generate a net lepton
number, which is then processed by the anomolous $B + L$ violation in
the \sm\ into a net baryon number.
The primary motivation for this extension lies in the fact that it
provides, via the see-saw mechanism, 
a framework for understanding the smallness of the
left-handed neutrino masses suggested by the atmospheric and solar
neutrino data. 

\section{Baryogenesis in the MSSM}

The squarks (scalar partners of the quarks) present in the MSSM get
contributions to their masses from supersymmetry breaking, as well as
from electroweak symmetry breaking via the Higgs mechanism. In
particular, $\tilde t_L$ and $\tilde t_R$ the scalar partners of the
top quark get a large contribution from the Higgs mechanism due to the 
size of the top quark Yukawa coupling to the Higgs boson. 
If the supersymmetry breaking mass of the $\tilde t_R$ is negligible,
and there is no \tl - \tr\ mixing, 
(the $\tilde t_R$ is chosen to be light in order to avoid conflicts
with the $\rho$ parameter if the $\tilde t_L$ were light),
Eq.~(\ref{smtransition}) gets modified to
\beq
\frac{H(T_0)}{T_0} \sim \frac{2 M_W^3 + M_Z^3 + 2 m_t^3}{2 m_H^2 v}  
                   \sim 3
\eeq
for $m_t = 175$ GeV. Thus we see, 
the condition of Eq. (\ref{firstorder}) can be satisfied and we have 
a strongly first order phase transition. 
This simple relation is modified by the presence of supersymmetry
breaking masses, $\tilde t_L - \tilde t_R$ mixing, and finite
temperature effects.
A detailed analysis \cite{SUSY_PT} shows that an electroweak phase 
transition strong enough to allow baryogenesis is possible if
$
m_{\tilde t_R} \le 175$ GeV and $m_H \le 115$ GeV.
Moreover, efficient baryogenesis requires rapid intraconversion
between the particles and their supersymmetric partners. This means
that most of the supersymmetric particles and especially the gauginos
(fermionic partners of the gauge bosons) 
must also have masses of order $T_0 \sim 100$ GeV, where $T_0$ is the
critical temperature for the electroweak phase transition.
Besides the obvious direct search implications of these
light sparticles and Higgs boson, the light \tr\ and charginos 
also result in large contributions to $B-\bar B$ mixing 
This is
because the $b_L-\tilde t_R -\tilde h$ coupling, proportional to the
top quark mass, removes the possibility of any GIM cancellation 
of its contribution \cite{worah1}.

The most effective way to generate a particle number 
asymmetry for some species is to arrange that, during the electroweak
phase transition, a $CP$ violating space-time dependent phase 
appears in the mass matrix for that species. If this
phase cannot be rotated away at subsequent points by the same unitary
transformation, it leads to different propagation probabilities
for particles and anti-particles, thus resulting in a particle number
asymmetry. The existence of two Higgs fields in the MSSM 
makes this possible. If $\tan\beta$ (the ratio of the expectation
values of the two Higgs fields)
changes as one traverses the bubble wall separating 
the symmetric phase from the broken one,
particle number asymmetries can be generated, which 
will be proportional to $\Delta\beta$, the change in $\beta$
across the bubble wall \cite{huet-nelson}.

It has been estimated that 
$\Delta\beta \propto m_h^2/m_A^2 \sim 0.01$ for the pseudoscalar Higgs
boson mass $m_A = 200-300$ GeV \cite{carena2}. 
This can actually be turned into an upper bound for $\Delta \beta$ 
using the
relation $m_{h_+}^2=m_A^2+m_W^2$, where $m_{h_+}$ is the charged Higgs
boson mass. Charged Higgs bosons make large positive contributions
to the $b \to s \gamma$ decay rate. 
The current experimental value
for $Br(b \to s \gamma)$ already sets the limit $m_{h_+} \ga 300$ GeV
at the $2\sigma$ level \cite{misiak}. This then implies $\Delta\beta \la
0.01$ through the relations above.

Baryogenesis in the MSSM proceeds most efficiently through the
generation of higgsino number or axial squark number in the bubble
wall, which then diffuses to the symmetric phase. Here, they bias the
\sm\ $B+L$ violation to produce a net baryon number
\cite{huet-nelson,carena2,worah}.
In this paper we present the special case of baryogenesis through the
production of axial squark number, where the CKM phase responsible for 
kaon $CP$ violation is also directly responsible for baryogenesis
\cite{worah}. 

Consider the mass squared matrix for the up-type squarks:
\beq
M^2_{\tilde u} = \left(\begin{array}{cc}
                 M^2_{\tilde u_{LL}} & M^2_{\tilde u_{LR}} \\
		M^{2\dagger}_{\tilde u_{LR}} & M^2_{\tilde u_{RR}}
		  \end{array} \right)
\eeq
where 
\beqa
M^2_{\tilde u_{LL}}&=&m^2_{Q}A_{U_{LL}}+(F,~D)~{\rm terms}, \nonumber \\
M^2_{\tilde u_{RR}}&=&m^2_{U}A_{U_{RR}}+(F,~D)~{\rm terms}, \nonumber \\
M^2_{\tilde u_{LR}}&=&m_Av_2\lambda_UA_{U_{LR}}+\mu v_1\lambda_U.
\eeqa
where $M_Q$, $M_U$, and $M_A$ are supersymmetry breaking masses
$\lambda_U$ is the Yukawa coupling matrix for up-type quarks, and the
$A_U$'s are dimensionless matrices. 
Concentrating only on the production of \tr, and using 
$m_{\tilde t_R}=175$ GeV, $m_{\tilde t_L}=300$ GeV, 
and $\tan\beta \sim 1$ we obtain the result
\beq
\frac{n_B}{s} \simeq
                10^{-8}\frac{\kappa\Delta\beta}{v_w}\frac{m_A}{T_0}
                \frac{|\mu |}{T_0} Im[e^{i\phi_B}
                A^{\dagger}_{U_{LR}}\lambda^{\dagger}_U\lambda_U]_{(3,3)}
\label{bnumber}
\eeq
$\kappa$ is related to the rate of anomolous $B+L$ violation,
$\Gamma_{B+L}=\kappa \alpha_w^4 T$. There is a large uncertainty in its
precise value, with current estimates giving $\kappa = 1-0.03$
\cite{sphalerons}.
$v_w \simeq 0.1$ is the velocity of the wall separating the phase
where electroweak symmetry is broken (the Higgs field has an
expectation value) from where it is unbroken (the Higgs field has no
expectation value). $\Delta\beta \la 0.01$,
and $T_0 \sim m_A \sim |\mu| \sim 100$ GeV.
The approximations made in deriving Eq. (\ref{bnumber}) and their
validity our outlined in \cite{huet-nelson}. If \tl\ and \tr\ have very
different masses there is a suppression of the baryon asymmetry by
$m_{\tilde t_R}^2/m_{\tilde t_L}^2$ 
that is not explicit in their work. Thus the
estimate of Eq. (\ref{bnumber}) would be modified if 
$m_{\tilde t_L} \gg 300$ GeV. 

Consider the possibility that the supersymmetric
parameters $A_{U_{LR}}$ and $\mu$ are real, with all the $CP$ violation
being in the quark mass matrix \cite{worah}. Notice that
$\lambda^{\dagger}_U\lambda_U$ in Eq. (\ref{bnumber}) is Hermitian,
hence the phase is on one of the off-diagonal terms. 
One then requires $A_{U_{LR}}$ to have off diagonal entries in order
to move this phase to the (3,3) element of the product
$A^{\dagger}_{U_{LR}}\lambda^{\dagger}_U\lambda_U$.
These large off-diagonal terms in $A_{U_{LR}}$ 
always lead to large $D-\bar D$ mixing due to gluino mediated box 
diagrams. The magnitude of the mixing is generically about an order
of magnitude lower than the current experimental bound $\Delta(m_D) < 1.3
\times 10^{-13}$ GeV. Further, given the hierarchical
structure of the quark masses and mixings, 
one expects the largest off-diagonal entry in 
$\lambda^{\dagger}_U\lambda_U$ to be $\sim \theta_C^2 \sim 0.04$.
For example the ansatz $\lambda_U = V^{\dagger}_{CKM}\hat
{\lambda}_U V_{CKM}$ where $V_{CKM}$ is the CKM matrix, and $\hat
\lambda_U$ is the diagonal matrix of up-type Yukawa couplings can lead to
\beq
Im[A^{\dagger}_{U_{LR}}\lambda^{\dagger}_U\lambda_U]_{(3,3)} =
\lambda_t^2|V_{cb}|\sin\gamma 
\eeq
for
\beq
A_{U_{LR}}=\left(\begin{array}{ccc}
            1&0&0 \\
            0&1&1 \\
            0&1&1 \end{array}\right),
\eeq
where 
$\gamma \sim 1$ 
is the phase in the CKM matrix.
Thus, we see that the baryon asymmetry is directly
related to the phase responsible for $CP$ violation in $K-\bar K$
mixing. We can obtain 
a large enough baryon asymmetry [{\em cf} Eq. (\ref{bnumber})] for
$\kappa = 1,~\Delta\beta =0.01$.

\section{Baryogenesis via Leptogenesis}

The idea that one can obtain a baryon asymmetry by first generating a
lepton asymmetry was first proposed in \cite{FY}, and subsequently
explored in several papers \cite{others}. As mentioned earlier, $B + L$ 
is anomously violated in the \sm, and the rate for this 
process is large at high temperatures. However, $B - L$ is
conserved. Thus given enough time for the $B+L$ violating processes to 
act, we obtain the relations:
\beqa
(B - L)_f &=& (B - L)_i \nonumber \\
(B + L)_f &=& 0
\eeqa
where the subscripts $f$ and $i$ stand for final and initial
respectively. Thus if one started with zero initial baryon number, but 
non-zero initial lepton number one would obtain the final condition
$B_f = -L_i$ (this relationship is slightly modified by a careful
consideration of all the \sm\ interactions \cite{HT}. 
The initial lepton number asymmetry is obtained by the $CP$ and lepton
number violating decay of heavy right-handed Majorana neutrinos.

Consider a model with right-handed Majorana neutrinos $N_R$. By
definition these fields are self conjugate, $N_R^c = N_R$
where the superscript $c$ denotes the charge conjugated field.
Thus given the Yukawa interaction
\beq
{\cal L}_Y = - h_{ij}\bar l_L^i N_R^j H + h.c.
\label{yuk}
\eeq
where $h_{ij}$ is the matrix of Yukawa couplings, 
the $l_L$ are left-handed \sm\ leptons and $H$ is the Higgs
field one finds that $N_R$ can decay into both light leptons and
anti-leptons. If
these decays are $CP$ violating they will generate an excess of one over 
the other. 
Let us define an asymmetry 
\beq
\delta = \frac{\Gamma - \Gamma^{CP}}{\Gamma + \Gamma^{CP}}
\label{defdelta}
\eeq
where $\Gamma$ is the decay rate into leptons, and $\Gamma^{CP}$ into
anti-leptons.
In the case that the heavy neutrinos are not degenerate in mass, which 
is the case we study here, it
is sufficient to consider only $CP$ violation in the decays of the heavy 
neutrinos (direct $CP$ violation). One then obtains the result
\cite{FY} 
\beq
\delta = \frac{1}{2\pi(h^{\dagger}h)_{11}} 
         \sum_{j=1}^6 {\rm Im}[(h^{\dagger}h)_{1j}]^2 
                                      f(m_j^2/m_1^2),
\label{delta}
\eeq
where $f(x)$ is a kinematic function of order one for reasonable
choices of the masses \cite{FY}.
%
%\beq
%f(x) = \sqrt{x} \left[1-(1+x) {\rm ln}\left(\frac{1+x}{x}\right)\right]
%\eeq
%
The subscript 1 in the terms above is due to the fact that the lepton
asymmetry is generated by the decay of the lightest of the
right-handed neutrinos (any asymmetry generated by the heavier
right-handed neutrinos will be washed out by the decays of the
lightest). 

The first constraint on  
the mass scale of the $N_R$ is obtained by insisting that it be
out of thermal equilibrium with the rest of the universe when it
decays.
This will hold if it lives till the universe has cooled to a
temperature below the mass of the particle. This condition is encoded
in the requirement that
\beq
\Gamma_R  \le  H(T=m_R) 
\eeq
where $\Gamma_R$ is the decay rate of the right-handed neutrino with
mass $m_R$, and 
$H$ is the Hubble constant. This translates to
\beq
\frac{(h^{\dagger}h)_{11}m_R}{8\pi} \la \frac{20 m_R^2}{M_P}
\Rightarrow
\frac{(h^{\dagger}h)_{11}}{m_R} \la 10^{-16}~{\rm GeV}^{-1}
\label{hubble}
\eeq
if the dominant decay is via the Yukawa coupling of
Eq. (\ref{yuk}). 
The second constraint is obtained by insisting that the heavy Majorana 
mass scale explain the solar and atmospheric neutrino data. If we
assume that the observed deficit in $\nu_e$'s from the sun 
is due to $\nu_e - \nu_{\mu}$ mixing, 
then the mass squared difference 
$\Delta m^2 \sim 10^{-6} ~{\rm eV}^{2}$, preferred by the data, 
implies $m_{\nu_{\mu}} \sim 10^{-3}$ eV. Similarly,
assuming the deficit in atmospheric $\nu_{\mu}$'s is due to 
$\nu_{\mu} - \nu_{\tau}$ mixing, then the preferred mass squared
difference, $\Delta m^2 \sim 10^{-3} ~{\rm eV}^{2}$, implies 
$m_{\nu_{\tau}} \sim 3 \times 10^{-2}$ eV. 
The see-saw mass relations
\beq
m_{\nu_{\mu}} \sim \frac{m_{\mu}^2}{m_R}; ~~ 
m_{\nu_{\tau}} \sim \frac{m_{\tau}^2}{m_R}
\eeq
then imply that $m_R \sim 10^{10} - 10^{11}$ GeV. Eq. (\ref{hubble})
then tells us that $h_{11} \sim 10^{-2} - 10^{-3}$ (assuming a
hierarchical matrix of Yukawa couplings). 
Note, that if the electron gets its mass at tree level from the Yukawa 
coupling $h$ as in the \sm, one would obtain $m_e = h_{11}v = 1$ GeV,
for $v=246$ GeV, which is too large by several orders of magnitude. In 
order for this model to work, one has to impose symmetries such that
the \sm\ fermions only get their masses at the loop level. In such a
case the fermion masses would be proportional to the squares of the
Yukawa coupling constants, and one obtains 
$m_e = h_{11}^2 v = 0.2$ MeV for $h_{11} = 10^{-3}$ which is the
correct order of magnitude. 
It is indeed possible to construct a model that incorporates all these 
requirements \cite{me2}. 
Moreover, in this model, the $CP$ violation responsible 
for baryogenesis is related in a calculable way to the $CP$ violation
present in the CKM matrix.

The model is based on the 
$SU(4) \times SU(2)_L \times SU(2)_R$ group.
The \sm\ fermions transform in the usual
representations:
\beq
\Psi_L^i\sim   ({4},{2},{1})^i\equiv
             \left(\begin{array}{cccc} u_1& u_2& u_3& \nu \\
                           d_1& d_2& d_3& e^- \end{array}\right)_L^i
\eeq
\beq
\Psi_R^i\sim   ({4},{1},{2})^i\equiv
             \left(\begin{array}{cccc} u_1& u_2& u_3& N \\
                           d_1& d_2& d_3& e^- \end{array}\right)_R^i
\eeq
where $i=1,2,3$ is a generation index, and we have included a right
handed neutrino $N$. We add to this three
generations of (right-handed) sterile neutrinos
\beq
s^i \sim  ({1},{ 1},{ 1})^i.
\eeq
The matter spectrum is supersymmetric, so the scalars 
$\tilde \Psi_L^i$, $\tilde\Psi_R^i$, and $\tilde s^i$ 
in the model transform in exactly the same way.
%
%\beq
%\tilde\Psi_L^i \sim   ({4},{2},{1})^i \equiv 
%          \left(\begin{array}{cccc} \tilde u_1& \tilde u_2& 
%                                    \tilde u_3& \tilde \nu \\
%                           \tilde d_1& \tilde d_2& 
%                           \tilde d_3& \tilde e^- \end{array}\right)^i,
%\eeq
%
%\beq
%R^i\sim   ({ 4},{ 1},{ 2})^i \equiv 
%          \left(\begin{array}{cccc} R_{u1}& R_{u2}& R_{u3}& R_{N} \\
%                           R_{d1}& R_{d2}& R_{d3}& R_{e} \end{array}\right)^i,
%\eeq
%
%and
%
%\beq
%\sigma^i \sim ({ 1},{ 1},{ 1})^i.
%\eeq
%
We will impose a discrete $Z_3$ symmetry on the gauge singlets 
(broken by the interactions of the \sm\ particles) under
which $s^j \rightarrow e^{-i(j\pi)/3} s^j$ and $\tilde s^j \rightarrow
e^{i(2j\pi)/3}\tilde s^j$. This permits us to make the Lagrangian $CP$
invariant, with the vacuum expectation values of the $\tilde s^j$
breaking $CP$ spontaneously.
We can choose parameters for the scalar potential such that it
is minimized when
\beq
\langle\tilde s\rangle_j = \frac{v_0}{\sqrt 2}e^{i\alpha_j};~~~
\langle \tilde N\rangle_j = \frac{v_R}{\sqrt 2}\delta_{1j};~~~
\langle \tilde \nu\rangle_j = \frac{v_L}{\sqrt 2}\delta_{1j}
\label{vevs}
\eeq
with $|v_0| > |v_R| \gg |v_L|$. This provides the correct pattern of
symmetry breaking. 

The Yukawa interactions are given by 
\beq
{\cal L_{Y}} = -y_i(\bar s^c)^i s^i \tilde s^i  
               -(\kappa_L^a)_{ij}\bar \Psi_L^i s^j \tilde\Psi_L^a
               -(\kappa_R^a)^T_{ij}\bar \Psi_R^i (s^c)^j \tilde\Psi_R^a + 
                {\rm h.c.},
\label{gaugeyuk}
\eeq
with all of the coupling constants real. 
However, the mass matrix of the $s_i$ will contain the phases $\alpha$ 
due to the spontaneous breaking of $CP$ invariance when the $\tilde s_i$ 
obtain a vacuum expectation value.
Note that since $\bar \Psi_L \Psi_R$ transforms as $(1,2,2)$ and there are
no scalars in this representation, none of the \sm\ fermions get
masses at tree level. Their masses are generated at one loop by
diagrams involving the $(s_i)$ on the internal lines. The $CP$
violating phases in the quark mass matrices and hence in the CKM
matrix are a function of the phases $\alpha$ in the masses of the $s_i$.

The out of equilibrium decays 
of the $s_i$ generate the lepton (and hence baryon) asymmetry.
It is this same phase $\alpha$ that is responsible for the
$CP$ violation in these decays. 
Thus one obtains a relationship between the
CKM phase and the phase responsible for the baryon asymmetry. 

\section{Conclusions}

We have presented an overview of two models of baryogenesis that also
have other low energy experimental consequences. 
Baryogenesis in the MSSM is possible if the Higgs and \tr\ are
light. Moreover, one expects large contributions to the $b \to s
\gamma$ rate, and the $B-\bar B$ mixing amplitude. Baryogenesis via
the decay of heavy neutrinos can be constrained by insisting that they
be at the see-saw scale implied by the solar and atmospheric neutrino
data. We have presented
specific implementations of these models where the CKM phase
responsible for $CP$ violation in the neutral kaons is related to the
phase responsible for the baryogenesis. 

{\em Acknowldegements}
This work was supported in part by the National Science Foundation
under grant PHY-95-147947 and 
by the U.S. Department of Energy under Contract DE-AC03-76SF00098.

%\newpage

\end{document}